\documentclass[aps,prl,twocolumn,amsfonts,amssymb,amsmath,floats,floatfix,showpacs,preprintnumbers,superscriptaddress]{revtex4-1}

\usepackage[mathscr]{eucal}
\usepackage{graphicx}
\usepackage{hhline}
\usepackage{psfrag}
\usepackage{enumerate}
\makeatletter
\@addtoreset{equation}{section}
\makeatother

\usepackage{color}
\usepackage{dcolumn}
\usepackage{bm}
\usepackage{hyperref}
\usepackage{relsize}

\def\be{\begin{equation}}
\def\ee{\end{equation}}
\def\bdm{\begin{displaymath}}
\def\edm{\end{displaymath}}
\def\bea{\begin{eqnarray}}
\def\eea{\end{eqnarray}}

\def\ri{{\rm i}}

\def\XXint#1#2#3{{\setbox0=\hbox{$#1{#2#3}{\int}$}
    \vcenter{\hbox{$#2#3$}}\kern-.5\wd0}}

\def\be{\begin{equation}}
\def\ee{\end{equation}}

\def\beq{\begin{equation}}
\def\eeq{\end{equation}}

\newcommand{\rd}{\mbox{d}}
\newcommand{\re}{\mbox{e}}

\begin{document}

\title{ From Fuchsian differential equations to integrable QFT}

\author{ V.V. Bazhanov}
\affiliation{Department of Theoretical Physics,
         Research School of Physics and Engineering,
    Australian National University, Canberra, ACT 0200, Australia}
\affiliation{Mathematical Sciences Institute,\\
      Australian National University, Canberra, ACT 0200,
      Australia}
\author{ S.L. Lukyanov}
\affiliation{NHETC, Department of Physics and Astronomy
     Rutgers University
     Piscataway, NJ 08855-0849, USA}
\affiliation{L.D. Landau Institute for Theoretical Physics\\
  Chernogolovka, 142432, Russia}

\begin{abstract}
We establish an intriguing correspondence between 
a special set of classical solutions of  the
modified sinh-Gordon equation (i.e., Hitchin's ``self-duality'' equations)
on a punctured Riemann sphere and a set of 
stationary states in the finite-volume Hilbert space of the
integrable 2D  QFT  introduced by V.A. Fateev. 
Potential applications of this correspondence to the problem
of non-perturbative quantization of classically integrable non-linear
sigma models are briefly discussed.
A detailed account of the results announced in this letter 
is contained in a separate publication [arXiv:1310.4390].


\end{abstract}

\maketitle

It is difficult to assign a precise mathematical meaning for the
concept of integrability in Quantum Field Theory.  A naive intuition
goes back to Liouville of the $19^{\rm th}$ century and suggests an
existence of a sufficiently large set of mutually commuting operators
whose joint spectra fully specify stationary states of the quantum
system.  For deeper insights, it is useful to consider 2D Conformal
Field Theory (CFT), where significant simplifications occur due to the
presence of an infinite dimensional algebra of (extended) conformal
symmetry \cite{Zamolodchikov:1989zs}. 
For a finite-size 2D CFT (with
the spatial coordinate compactified on a circle of the circumference
$R$), a mathematically satisfactory construction of an infinite set of
mutually commuting local Integrals of Motion (IM) can be given and the
simultaneous diagonalization of these operators turns out to be a
well-defined problem within  the representation theory of the associated
conformal algebra.

Different conformal algebras, as well as different sets of mutually
commuting local IM yield a variety of integrable structures in CFT.
The series of works
\cite{BLZ} was dedicated
to the simplest of these structures, associated with the
diagonalization of the local IM from the quantum KdV hierarchy
\cite{KDV}.  Subsequent
studies of this problem culminated in a rather surprising
link between the integrable structures of CFT and spectral
theory of Ordinary Differential Equations (ODE) 
\cite{ODE/IM,Bazhanov:2003ni}.  In particular,
in \cite{Bazhanov:2003ni} a one-to-one correspondence was conjectured
between the joint eigenbasis of the IM from the quantum KdV hierarchy
and a certain class of differential operators of the second order
$-\partial_z^2+V_L(z)$, with singular potentials
$V_L(z)$ (``monster'' potentials in terminology of
\cite{Bazhanov:2003ni}). Apart from a regular singularity at $z=0$ and
an irregular singular point at $z=\infty$, the monster potentials
possess $L$ regular singular points $\{x_a\}_{a=1}^L$. These
potentials are not of much immediate interest in quantum mechanics,
but arise rather naturally in the context of the theory of
isomonodromic deformations.  Solutions of the corresponding
Schr\"odinger equations are single valued (monodromy-free) at $z=x_a$
and their monodromy properties turn out to be similar to those of the
radial wave functions for the three-dimensional isotropic anharmonic
oscillator.  The monodromy-free condition was formulated in a form of
a system of $L$ algebraic equations imposed on the set
$\{x_a\}_{a=1}^L$.
The correspondence  
proposed in \cite{Bazhanov:2003ni} precisely relates
the set of monster potentials $V_L(z)$ and the joint eigenbasis for all
quantum KdV integrals of motion in the level $L$ subspace of the highest weight
representation of the Virasoro algebra. In particular, 
this implies that a number of the potentials 
$V_L(z)$ with a given value of $L$ exactly coincides with a number of
partitions ${\tt p}_1(L)$ of the integer $L$ into parts of one kind.

Since 1998, the link to the  spectral theory of ODE have been
extended to a large variety of  integrable CFT structures (for
a review, see \cite{Dorey:2007zx}), so that a natural question has
emerged on whether a similar  relation exist for {\em massive} integrable
QFT. This question  remained more or less dormant until 
the work \cite{Gaiotto:2008cd}, after which 
the so-called thermodynamic Bethe
Ansatz equations have started to appear in different contexts of SUSY gauge
theories
\cite{TBA}.
These remarkable developments have led to the work
\cite{Lukyanov:2010rn}, which  
established a link between the eigenvalues of IM in the  vacuum
sector of  the massive sine/sinh-Gordon model and some new spectral
problem generalizing the one from \cite{ODE/IM}. 

This work is aimed to extend the results of 
\cite{Bazhanov:2003ni,Lukyanov:2010rn}  
and provide an explicit example of the correspondence 
between an infinite set of 
stationary states of massive integrable QFT in a finite volume and
a set of singular  differential operators of a certain type.
At first glance, the best candidate for such study should be 
the sine-Gordon model,
which always served as a basis for the development of integrable QFT.
However, in spite of some technical complexity, a more general model
introduced by Fateev \cite{Fateev:1996ea} (which contains
the sine-Gordon model as a particular case) turned out to be more
appropriate for this task.
The situation here is analogous to that in the  Painlev\'e  theory.
Even though the Painlev\'e VI is the most complicated
and general  equation  in the  Painlev\'e classification, 
geometric structures behind this  equation are much more transparent 
than those related to its degenerations. From this point of view, the fact that
the sine-Gordon model is a certain degeneration of the Fateev model,
could be understood as a QFT version of the relationship between  the
Painlev\'e VI and a particular case of
Painlev\'e III. 

Our  starting point   is a special class of Fuchsian differential operators of the second order
${\cal D}=-\partial_z^2+T_L(z)$
with $3+L$ regular singular points at
$z=z_1,z_2,z_3$ and $z=x_1,\ldots, x_L$.
The variable $z$ can be regarded as a complex coordinate on the Riemann sphere with $3+L$ punctures.
Projective transformations of $z$ allows one to send three points
$z_i$ to any designated positions. 
At the same time other parameters of $T_L(z)$ are chosen in such a way that the
remaining $L$  regular singular points satisfy the monodromy-free
condition. Therefore, monodromy  properties of the differential operator ${\cal D}$ with $L>0$ 
turn out to be similar to those for $L=0$ (i.e. the ordinary
hypergeometric  differential operator of the second order). 

Next, we consider more general differential operators of the form 
${\cal D}{(\lambda)}=-\partial_z^2
+T_L(z)+\lambda^2\ {\cal P}(z)$, where 
$${\cal P}(z)=\frac{(z_3-z_2)^{a_1}\,(z_1-z_3)^{a_2}\,(z_2-z_1)^{a_3}}
{(z-z_1)^{2-a_1}(z-z_2)^{2-a_2}(z-z_3)^{2-a_3}}
$$
and 
parameters $0<a_i<2$ obey the constraint $a_1+a_2+a_3=2$.
Due to the last relation, ${\cal P}(z)(\rd z)^2$ transforms as a
quadratic differential under $\mathbb{PSL}(2,\mathbb{C})$
transformations and the punctures  $z_1,z_2,z_3$ on the Riemann sphere
can still  be sent to any desirable positions.
The monodromy properties of ${\cal D}(\lambda)$ for $\lambda\not=0$ change
dramatically  
in comparison with the case $\lambda=0$. However,  one can still find  
positions of the punctures $x_1,\ldots, x_L$ so that they remain 
monodromy-free singular points   for any values of $\lambda$. In this
case the coordinates $\{x_i\}_{i=1}^L$ obey a the system 
of $L$ algebraic equations 
similar to that from \cite{Bazhanov:2003ni},
and the  moduli space of the operators ${\cal D}{(\lambda)}$ constitute a 
{\it finite discrete subset} ${\cal A}^{(L)}$ of the moduli space of 
${\cal D}{(0)}=-\partial_z^2+T_L(z)$ \cite{foot1}.
It appears that, for a given $L$, the cardinality  of ${\cal A}^{(L)}$
coincides with the number of partitions ${\tt p}_3(L)$ of the integer
$L$ into parts of three kinds. 
We interpret this fact in the spirit of
\cite{Bazhanov:2003ni}, and present arguments in support of the existence 
of 
a one-to-one correspondence 
between  the elements of ${\cal A}^{(L)}$
and  the level-$L$  common eigenbasis 
of  local   IM  of  the integrable hierarchy 
introduced by Fateev in \cite{Fateev:1996ea} 
(see ref.\cite{BL2013} for details).
The  arguments   closely follow  the line of 
\cite{BLZ}
adapted to the algebra of extended conformal symmetry, which can be
regarded as a quantum 
Hamiltonian reduction of the exceptional affine 
superalgebra ${\hat D}(2,1;\alpha)$ \cite{Feigin:2001yq} 
(the ``corner-brane'' $W$-algebra, in terminology of \cite{Lukyanov:2012wq}).

The above structure can be generalized 
to the case of massive QFT. The construction is based on the idea  
from \cite{Lukyanov:2010rn}, which was inspired by the works
\cite{Gaiotto:2008cd,TBA}.
As far as our attention has been restricted to the case of CFT, there was
no need to separately consider  
the antiholomorphic  differential operator
${\bar {\cal D}}{({\bar \lambda})}=
-\partial_{\bar z}^2+{\bar T}_{\bar L}({\bar z})+{\bar \lambda}^2\ {\bar {\cal P}}({\bar z})$, 
since there is only a nomenclature 
difference between the holomorphic and antiholomorphic cases. In massive QFT
one should substitute $({\cal D}{(\lambda)},\,{\bar
  {\cal D}}{({\bar \lambda})})$
by a pair of $(2\times 2)$-matrix  valued differential operators
$\big({\boldsymbol { D}}{(\lambda)},{\bar {\boldsymbol   D}}{({\bar
    \lambda})}\big)= 
\big(\partial_z-{\boldsymbol
  A}_z,
\partial_{\bar z}-{ {\boldsymbol A}}_{\bar z}\big)$
with
\bea{\boldsymbol
  A}_z&=&-{\textstyle\frac{1}{2}}\ \partial_z\eta\,\sigma_3+
\sigma_+\,\re^{+\eta}+
\sigma_-\, \lambda^2\, {\cal P}(z)\, \re^{-\eta}\nonumber\\
{ {\boldsymbol A}}_{\bar z}&=&+
{\textstyle\frac{1}{2}}\ \partial_{\bar
  z}\eta\,\sigma_3+\sigma_-\, 
\re^{+\eta}+\sigma_+\,\bar{\lambda}^2\,  {\bar {\cal P}}({\bar z})\,\re^{-\eta}\ .\nonumber
\eea
where $\sigma_3,\sigma_\pm=(\sigma_1\pm \ri \sigma_2)/2$ are the 
standard Pauli matrices. 
The pair $({\boldsymbol A}_z,\,  {\boldsymbol A}_{\bar z})$ forms an
$\mathfrak{sl}(2)$  connection whose flatness 
is a necessary condition for the existence of solution of the
auxiliary linear problem 
${\boldsymbol { D}}{(\lambda)}\, {\boldsymbol \Psi}={\bar {\boldsymbol  D}}{({\bar \lambda})}\,{\boldsymbol \Psi}=0.
$
The zero-curvature condition yields
the Modified Sinh-Gordon equation (MShG):
$$
\partial_z\partial_{\bar z}\eta-\re^{2\eta}+ \rho^4\ {\cal P}(z){\bar {\cal P}}({\bar z})\, \re^{-2\eta}=0\ \ \ 
\ \ \ \ \ \ \ (\rho^2={\lambda{\bar\lambda}})\ .
$$
We consider a particular class of singular solutions of this equation,
defined of the following requirements:
\begin{enumerate}[(i)]
\item
$\re^{-\eta}$ should be a {\it  smooth, single valued
  complex function without zeroes} 
on the Riemann sphere with $3+L+{\bar L}$
punctures. 
Since
$z=\infty$ is assumed to be a regular point on the  sphere, then
$\re^{-\eta}\sim |z|^{2}$ as $|z|\to  \infty$.

\item
$\re^{-\eta}$ develops a singular behavior at $z=z_i$:
$\re^{-\eta}\sim |z-z_i|^{-2m_i}$ as $ |z-z_i|\to  0\ \ (i=1,2,3)$, 
where  parameters $m_i$ are  restricted within  the domains
$-{\textstyle\frac{1}{2}}\leq m_i\leq -{\textstyle\frac{1}{4}}\ (2-a_i)$ \cite{foot2}.

\item\label{mfree}
$\re^{-\eta}$ also develops a singular behavior  at
$z=x_a\ (a=1,\ldots, L)$
and ${\bar z}={\bar y}_b\ (b=1,\ldots, {\bar L})$:\ 
$\re^{-\eta}\sim\frac{{\bar z}-{\bar
    x}_a}{{z-x_a}},\ \ 
\re^{-\eta}\sim \frac{{ z}-{ y}_b}{{\bar z}-{\bar
    y}_b} .
$
The positions of these punctures 
are  constrained by the requirement
that $ \re^{\pm \frac{1}{2}\eta\sigma_3}\ {\boldsymbol \Psi}$ 
is single-valued in the neighborhood of the punctures
 $z=x_a$  and ${\bar z}={\bar y}_b$ 
(where ${\boldsymbol \Psi}$ is a general solution 
of the MShG auxiliary linear problem). 
\end{enumerate}

Following \cite{Feigin:2007mr}, the
above monodromy-free condition (\ref{mfree})
can be transformed into a set
of $L+{\bar L}$ constraints imposed on the 
regular part of local  expansions of $(\partial_z\eta,\partial_{\bar
  z}\eta)$  at  the monodromy-free punctures: 
$$\partial_z\eta=\frac{1}{z-x_a}+\frac{1}{2}\ \gamma_a+o(1)\, ,\ 
\partial_{\bar z}\eta=-\frac{1}{{\bar z}-{\bar
    x}_a}+o(1)$$
$$
\partial_{\bar z}\eta=\frac{1}{{\bar z}-{\bar y}_b}+\frac{1}{2}\ {\bar \gamma}_b+o(1)\, ,\ 
\partial_z\eta=-\frac{1}{z-y_b}+o(1)\, ,$$
where $\gamma_a=\partial_z\log {\cal P}(z)|_{z=x_a},\ {\bar \gamma}_b=
\partial_{\bar z}\log {\bar {\cal P}}({\bar z})|_{{\bar z}={\bar y}_b}$.

It is  expected that as far as 
positions of the  punctures $z_i$ are fixed, the triple ${\bf
  m}=(m_1,m_2,m_3)$ and 
the pair  $(L,{\bar L})$ are chosen,
the MShG equation has a {\it finite  set} ${\cal A}^{(L,{\bar
    L})}_{\bf m}$ of solutions 
satisfying  all  of the above requirements. 
Then we can define the moduli space ${\cal A}_{\bf m}$ which is a 
union of these  finite sets:
${\cal A}_{\bf m}=\cup_{L,{\bar L}}^\infty\,{\cal A}^{(L,{\bar L})}_{\bf m}$.
Notice that, to a certain extent, ${\cal A}_{\bf m}$ can be regarded
as a Hitchin moduli space \cite{Hitchin:1986vp}. 

An essential ingredient of the
formal theory of the MShG equation
is an existence of an infinite  hierarchy of  one-forms, 
which are closed by virtue of the MShG equation  itself.
With these forms, on can define an infinite set of conserved charges  
$\{{\mathfrak q}_{2n-1},\, {\bar {\mathfrak q}}_{2n-1}\}_{n=1}^\infty$, 
which can be used to characterize the elements of
the moduli space  ${\cal A}_{\bf m}$. 
Indeed, in spite of the fact that
the flat  connection 
${\boldsymbol A}={\boldsymbol A}_z\,\rd z+{\boldsymbol A}_{\bar z}\,\rd {\bar z}$
associated with an  element of ${\cal A}_{\bf m}$
is not single-valued on the punctured sphere,
it does return to the original value
after a continuation along the Pochhammer loop --- 
the contour $\gamma_P$ depicted in Fig.\ref{fig1av}:
\begin{figure}[!ht]
\centering
\includegraphics[width=5.  cm]{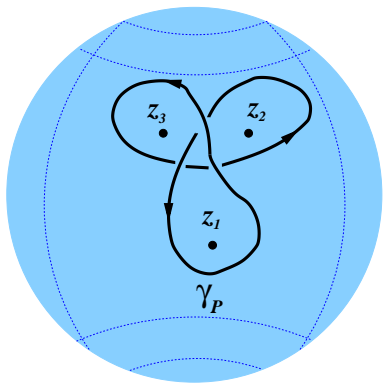}
\caption{The Pochhammer  loop on the  Riemann sphere.}
\label{fig1av}
\end{figure}

\noindent
Therefore one  can consider the Wilson loop
$W={\rm Tr}\big[{\cal P}\exp\big(\oint_{\gamma_P} {\boldsymbol  A}\big)\big]$,
whose significant advantage is
that it does not depend on the precise
shape of the integration contour. In particular, it
is not sensitive to deformations of $\gamma_P$
which sweep through
the monodromy-free punctures.
The Wilson loop is
an entire periodic function of the spectral parameter
$\theta=\frac{1}{2}\ \log(\lambda/{\bar\lambda})$, which 
possesses an asymptotic expansion:
\bea
\log W\asymp
\begin{cases}
   C\, \re^{\theta}+\sum_{n=1}^\infty   {\mathfrak
     q}_{2n-1}\ \re^{-(2n-1)\theta}\ \ \ 

(\theta\to+\infty)                          
\\
  C\, \re^{-\theta}+\sum_{n=1}^\infty  {\bar {\mathfrak
      q}}_{2n-1}\ \re^{(2n-1)\theta}\ \ \  
(\theta\to-\infty)
\end{cases}
\nonumber
\eea
Here  $C$ is some constant, whereas 
$\{{\mathfrak q}_{2n-1},\, {\bar {\mathfrak q}}_{2n-1}\}_{n=1}^\infty$
denotes the infinite set of conserved charges.

The main objective of this work  is to
propose the correspondence  between elements of 
the moduli space  ${\cal A}_{\bf m}$ 
and a subset ${\cal H}^{(0)}_{\bf k}$ of stationary states
of the Fateev model in a finite volume.
To describe ${\cal H}^{(0)}_{\bf k}$
explicitly, let us  recall some basic facts about this model. 

The Fateev model is governed by
the  following  Lagrangian in $1+1$ Minkowski
space
\bea
\nonumber
{\cal L}&=& \frac{1}{16\pi}\sum_{i=1}^3
\big( (\partial_t\varphi_i)^2-(\partial_x\varphi_i)^2\big)
+2\mu\ 
\big(\,\re^{\ri \alpha_3\varphi_3}\\
&\times&\cos(\alpha_1\varphi_1+\alpha_2\varphi_2)
+\re^{-\ri \alpha_3\varphi_3}\
\cos(\alpha_1\varphi_1-\alpha_2\varphi_2)\,\big)  .
\nonumber
\eea
for the three scalar fields $\varphi_i(x,t)$. 
Here $\alpha_i$ are  real   coupling constants subject to a single constraint
$\alpha_1^2+\alpha_2^2+\alpha_3^2=\frac{1}{2}$.
The parameter $\mu$ in the Lagrangian sets the mass scale, $\mu\sim [\,{\rm mass}\,]$.
We assume a finite-size geometry (with the
spatial coordinate $x$ compactified on a circle of
circumference $R$) and impose the periodic boundary conditions
$\varphi_i(x+R,t)=\varphi_i(x,t)$.
Due to the periodicity of the Lagrangian in $\varphi_i$,  
the space of states ${\cal H}$ in the model splits into orthogonal
subspaces ${\cal H}_{\bf k}$ characterized by the three  
``quasimomenta'' ${\bf k}=(k_1,k_2,k_3)$.
Similarly to the quantum mechanical problem of
a particle in a periodic potential,
the subspaces ${\cal H}_{\bf k}$ possess the band structure, 
where 
the subspace ${\cal H}^{(0)}_{\bf k}$, corresponding to  the first
Brillouin zone, is of primary interest. 
The Fateev model is integrable, in particular
it has an infinite set of commuting local integrals of motion 
$\mathbb{I}^{(+)}_{2n-1}$,\ $\mathbb{ I}^{(-)}_{2n-1}$, with 
$2n=2,\, 4,\,6,\,\ldots$
being the Lorentz spins of the associated local densities.
For
$0\leq k_i\leq\frac{1}{2}$, 
the sets of eigenvalues 
$\{I^{(+)}_{2n-1},I^{(-)}_{2n-1}\}_{n=1}^\infty$ fully specify the
common eigenbasis
of  the  local   IM in ${\cal H}^{(0)}_{\bf k}$.

In the recent paper \cite{LUK}  it was conjectured 
that  the set of {\it  vacuum} eigenvalues
$\{I^{(+)}_{2n-1},I^{(-)}_{2n-1}\}_{n=1}^\infty$
(i.e.  corresponding to the unique state  in ${\cal H}^{(0)}_{\bf k}$
with the lowest value of the energy $E=I^{(+)}_1+I^{(-)}_{1}$)
essentially 
coincides with the set of conserved charges
 $\{{\mathfrak q}_{2n-1},\, {\bar {\mathfrak q}}_{2n-1}\}_{n=1}^\infty$
associated with the unique element
${\cal A}^{(0,0)}_{\bf m}$ of  the moduli space  ${\cal A}_{\bf m}$,
provided that 
$$\alpha_i^2={\frac{a_i}{4}}\ ,\ \ \ \ \ \ \ \  k_i=\frac{1}{a_i}\ (2m_i+1)\ ,\ \ \ \ \ 
\ \ \ \mu R=2 \rho\ .
$$
In this work we extend the conjecture of \cite{LUK} to the 
whole  spectrum
of  the local IM in the subspace ${\cal H}^{(0)}_{\bf k}$.
Namely, we conjecture that the corresponding eigenvalues 
coincide with the values of the classical conserved charges
associated with the
elements of the moduli space ${\cal A}_{\bf m}$.
Thus, for the 
values of $k_i$, restricted to the segment $[\,0,\frac{1}{2}\,]$, 
there is a remarkable correspondence
between the joint  eigenbasis of the local IM in
${\cal H}^{(0)}_{\bf k}$ space and the solutions of MShG equation
specified by the elements of ${\cal A}_{\bf m}$.

Among various applications, the above 
correspondence between the classical and quantum integrable systems 
provides  a powerful tool for deriving  functional and integral equations
which determine
the spectrum of local IM in massive QFT.
We believe that this correspondence 
is, in fact, a general phenomenon which
open a new way of approaching integrable QFT and, most importantly, 
the problem of  non-perturbative quantization of classically integrable  nonlinear sigma models.
Here, we are motivated by the following consideration.

The above discussion has been focused on the ``symmetric'' regime 
of the Fateev model where
all the couplings   $\alpha_i$ are real, so that
the Lagrangian  is completely symmetric under simultaneous
permutations of the real fields  $\varphi_i$ and the real couplings $\alpha_i$.
The theory is apparently non-unitary in this case.
In the most interesting unitary regime
one of the couplings, say $\alpha_3$,
is pure imaginary: $\alpha_1^2>0,\  \alpha_2^2>0,\ \alpha_3^2:=-b^2<0$, 
and the Lagrangian becomes real
\bea
\nonumber
{\cal L}&=& \frac{1}{16\pi}\sum_{i=1}^3
\big( (\partial_t\varphi_i)^2-(\partial_x\varphi_i)^2\big)
-2\mu\ 
\big(\,\re^{b\varphi_3}\\
&\times&\cos(\alpha_1\varphi_1+\alpha_2\varphi_2)
+\re^{-b\varphi_3}\
\cos(\alpha_1\varphi_1-\alpha_2\varphi_2)\,\big)  .
\nonumber
\eea
The physical content in the unitary regime 
is  different from  the symmetric one.
However, assuming the same periodic boundary conditions  for each field $\varphi_i\  (i=1,2,3)$, 
we can  use the same symbols  ${\cal H}$ and ${\cal H}_{\bf k}$ 
to denote  the   spaces of states and their certain  linear subspaces
in the both cases (except that 
${\bf k}$ in the  unitary regime should be regarded 
as a pair of quasimomenta, ${\bf  k}=(k_1,k_2)$, because of lack
of periodicity in $\varphi_3$-direction).
As before, we focus on the component ${\cal H}^{(0)}_{\bf k}$
corresponding to the first Brillouin zone.
An important property of the local IM is that their existence and their  form
are not sensitive to the choice of parameter values.
Thus the eigenstates in  ${\cal H}^{(0)}_{\bf k}$ are again specified 
by the joint spectra of local IM. 

Having in mind   relations between ${\cal H}^{(0)}_{\bf k}$ and ${\cal A}_{\bf m}$ in the symmetric regime,  
let us consider the MShG equation in the regime $a_1,\, a_2>0,\,  a_3<0$
(the constraint $a_1+a_2+a_3=2$ is still assumed). 
A brief inspection shows that the set of requirements imposed on the MShG field
remains  quite  meaningful in this case. Only the asymptotic condition
which describes
the  behavior  of the solution in the vicinity of the third puncture
$z_3$, requires a special attention. 
For $a_i>0$ we had the freedom to control the asymptotic behavior of
$\eta$ as $z\to z_i$, with the arbitrary parameters $m_i$. If  
$a_3<0$, the situation is different: the leading asymptotic behavior
of the solution at $z=z_3$ is determined by the  
MShG equation itself \cite{Lukyanov:2010rn}:
$\re^{-\eta}\sim \big|{\cal P}(z)\big|^{-\frac{1}{2}}\propto |z-z_3|^{\frac{a_3}{2}-1}$. 
Taking this into account one can still define the 
moduli space ${\cal A}_{\bf m}$, which now will be labeled by the pair  ${\bf
  m}=(m_1,m_2)$.  
At the same time the definition of 
the set of the conserved charges
$\{{\mathfrak q}_{2n-1},\, {\bar {\mathfrak q}}_{2n-1}\}_{n=1}^\infty$ 
remains unchanged. 
We expect that  
the relation between the subspace ${\cal H}_{\bf k}^{(0)}$ and
the moduli space ${\cal A}_{\bf m}$ will continue to 
holds for the unitary regime as well.

The Fateev model in the unitary regime admits a dual description in
terms of the action  
$${\cal S}=\int\rd^2x\, G_{\mu\nu}(X)\, \partial_a X^\mu\partial_a X^\nu\ , $$
where
$G_{\mu\nu}$ is a certain two-parameter families
of metric on the topological three-sphere which possesses two $U(1)$
Killing vector fields \cite{Fateev:1996ea}. The sigma-model
description is especially useful in 
the strong coupling  limit
($\alpha_i^2,\ b^2\to \infty$ with  $\alpha_i^2/b^2$ kept fixed), which can be regarded as
the classical limit. Notice that  the classical  integrability of the theory 
was established only recently in  Ref.\cite{Lukyanov:2012zt}.

\bigskip
\noindent
{\bf Acknowledgments}: We are deeply indebted  to A.B. Zamolodchikov
for teaching us QFT and previous collaborations. 
The research of VB is 
partially supported by the Australian Research Council.


\begin{thebibliography}{99}
\bibitem{foot2}
At $m_i=
\frac{1}{2},\,\frac{1}{4}(2-a_i)$ the  leading  asymptotic  $\re^{-\eta}$ as $z\to z_i$  involves logarithms. 
Here we ignore such subtleties.
\end{thebibliography}
\end{document}